\documentclass{article}
\usepackage{dcase2024_techrep,amsmath,graphicx,url,times,booktabs, tabularx, lscape, pifont,cite}

\title{FMSG-JLESS Submission for DCASE 2024 Task4 on Sound Event Detection with Heterogeneous Training Dataset and Potentially Missing Labels}

\name{Yang Xiao$^{1}$,
      Han Yin$^{2}$,
      Jisheng Bai$^{2}$,
      and Rohan Kumar Das$^{1}$}

\address{$^1$Fortemedia Singapore, Singapore\\   
        $^2$Joint Laboratory of Environmental Sound Sensing, School of Marine Science and Technology,\\ Northwestern Polytechnical University, Xi’an, China\\
        \small\{xiaoyang, rohankd\}@fortemedia.com, \{yinhan, baijs\}@mail.nwpu.edu.cn
        }
\begin{document}

\ninept
\maketitle

\begin{sloppy}

\begin{abstract}
This report presents the systems developed and submitted by Fortemedia Singapore (FMSG) and Joint Laboratory of Environmental Sound Sensing (JLESS) for DCASE 2024 Task 4. The task focuses on recognizing event classes and their time boundaries, given that multiple events can be present and may overlap in an audio recording. The novelty this year is a dataset with two sources, making it challenging to achieve good performance without knowing the source of the audio clips during evaluation. To address this, we propose a sound event detection method using domain generalization. Our approach integrates features from bidirectional encoder representations from audio transformers and a convolutional recurrent neural network. We focus on three main strategies to improve our method. First, we apply mixstyle to the frequency dimension to adapt the mel-spectrograms from different domains. Second, we consider training loss of our model specific to each datasets for their corresponding classes. This independent learning framework helps the model extract domain-specific features effectively. Lastly, we use the sound event bounding boxes method for post-processing. Our proposed method shows superior macro-average pAUC and polyphonic SED score performance on the DCASE 2024 Challenge Task 4 validation dataset and public evaluation dataset.
\end{abstract}


\section{Introduction}
\label{sec:intro}
Sound event detection (SED)~\cite{sed,sed2, home,smarthome,smarthome2} involves identifying sound events from acoustic signals and accurately classifying them into specific categories with timestamps, considering various acoustic environments. DCASE 2024 Task 4~\cite{dcase2024}, entitled ``Sound Event Detection with Heterogeneous Training Dataset and Potentially Missing Labels," focuses on SED. This task follows up on DCASE 2023 Task 4A~\cite{fmsg1,fmsg2} and Task 4B~\cite{jless1} on the following aspects:
\begin{itemize}
    \item DCASE 2023 Task 4A evaluated systems for detecting sound events using weakly labeled data (without timestamps) and unlabeled data in the DESED dataset~\cite{desed}. The goal was to provide both event class and event time localization, despite multiple overlapping events in an audio recording.
    \item DCASE 2023 Task 4B evaluated systems for detecting sound events with soft labeled data from the MAESTRO dataset~\cite{maesro}. This task focused on exploring the significance of using soft labels for SED.
\end{itemize}

DCASE 2024 Task 4 aims to unify the setups of both the tasks of 2023 edition. Specifically, instead of training an SED model on each subtask separately with its dataset, an intriguing approach is to just train a single model on all available datasets. The goal is still to provide event classes along with their time boundaries, even with multiple overlapping events. This task explores leveraging training data with varying annotation granularity (temporal resolution, soft/hard labels). Systems will be evaluated on labels with different granularity to understand their behavior and robustness for various applications. Target classes in different datasets also differ, so sound labels present in one dataset might not be annotated in another. The systems need to handle potentially missing target labels during training and perform without knowing the origin of the audio clips at evaluation time. 

Although previous years' challenges, like the frequency dynamic convolutional recurrent neural network (FDY-CRNN)~\cite{fdy,fmsg4,ssp}, have shown notable performance in DCASE Task 4, this year's Task 4 introduces new challenges. The main challenge is how to combine heterogeneous training datasets from diverse domains with different annotations to improve performance. Deep neural networks struggle to generalize across diverse domains, leading to poor results in real-world scenarios. Therefore, domain generalization (DG)~\cite{dg} has become an essential research topic in fields like computer vision, audio processing, and natural language processing. Inspired by the exploration in DCASE Task 1~\cite{freqmixstyle}, which dealt with audio clips from multiple devices, we propose using the domain generalization approach for this year's Task 4.

In this technical report, we outline our contributions to our submission for DCASE 2024 Task 4. The primary contributions of our submissions are as follows:
\begin{itemize}
    \item We utilize the frame-level embeddings generated by the pretrained BEATs model in late-fusion with the FDY-CRNN and then fed into the recurrent neural network with the classifier.
    \item We leverage the DG to explore the appropriate way to use the heterogeneous training datasets from diverse domains. 
    \item We modify the baseline framework to independently compute the training loss for our model, which is specific to each dataset for their corresponding classes.
    \item We employ the sound event bounding boxes method as a post-processing method to further enhance the performance in the DESED dataset.
\end{itemize}

\section{Dataset}
\label{ssec:dataset}
The DCASE 2024 Challenge Task 4 comprises two datasets, and participants must use both in the training phase and provide one individual model that performs well for the two datasets.

\subsection{DESED dataset}

DESED~\cite{desed} consists of 10-second audio clips either recorded in a domestic environment or synthesized to reproduce such an environment. It features annotated sound events from 10 classes: alarm bell ringing, blender, cat, dishes, dog, electric shaver/toothbrush, frying, running water, speech, and vacuum cleaner. The synthetic part of the dataset is generated with Scaper~\cite{scaper} using foreground events from the Freesound datasets and backgrounds from YouTube videos and the Freesound subset of the MUSAN~\cite{musan} dataset. The real-world recording part comes from AudioSet~\cite{audioset} and includes a weakly annotated set (1,578 clips), an unlabeled set (14,412 clips), and a strongly annotated portion (3,470 clips).

\subsection{MAESTRO Real dataset}

MAESTRO Real~\cite{maesro, meestro1}, used in the DCASE 2023 Task 4B challenge, consists of a development set (6,426 clips) and an evaluation set of long-form real-world recordings. This dataset contains multiple temporally strong annotated events with soft labels from 17 classes. However, in this challenge, only 11 classes are evaluated, as the other 6 do not have confidence levels over 0.5. These classes are: birds\_singing, car, people\_talking, footsteps, children\_voices, wind\_blowing, brakes\_squeaking, large\_vehicle, cutlery\_and\_dishes, metro\_approaching, and metro\_leaving. This data was annotated using crowdsourcing, where temporally-weak labeling is combined with a sliding window approach to determine events' temporal localization. In order to obtain the soft labels, annotations of multiple annotators are aggregated via MACE~\cite{mace}. The recordings are taken from the TUT Acoustic Scenes 2016 dataset~\cite{tut} and are between 3 to 5 minutes long.
\section{Proposed Approach}
\label{sec:method}
\subsection{Baseline}
\label{ssec:baseline}
The baseline system is inherited from previous DCASE Task 4 challenges~\cite{fmsg1,dcase2022} and consists of a CRNN~\cite{crnn} that uses self-supervised features from the pre-trained BEATs~\cite{beats} model. First, the CRNN has a convolutional neural network (CNN) encoder with 7 convolutional layers, batch normalization, gated linear units, and dropout, followed by a bi-directional gated recurrent unit (biGRU) layer. Then, BEATs features are concatenated with the CNN-extracted features before the biGRU layer. Average pooling is applied to the BEATs features to match the sequence length of the CNN encoder. Finally, Attention pooling is used to derive clip-wise and frame-wise posteriors. During training, the BEATs model remains frozen, and the mean-teacher framework~\cite{meanteacher,meanteacherbaseline1} is used to leverage unlabeled and weakly labeled data. The attention pooling mechanism in the baseline model uses the softmax function over classes. Before applying softmax, values for unlabeled classes (not in the current clip dataset) are masked to negative infinity.

As a preprocessing step, some DESED events are mapped to similar classes in MAESTRO. For example, in DESED, ``speech" is a super-class for ``people talking," ``children's voices," and ``announcements" in MAESTRO. ``Dishes" in DESED corresponds to ``cutlery and dishes" in MAESTRO, and ``dog" in DESED is a super-class for ``dog bark". This mapping ensures that when computing the loss for MAESTRO, the network output for similar classes in DESED is adjusted accordingly.

\begin{figure}[t]
\centering  
\includegraphics[width=0.64\columnwidth]{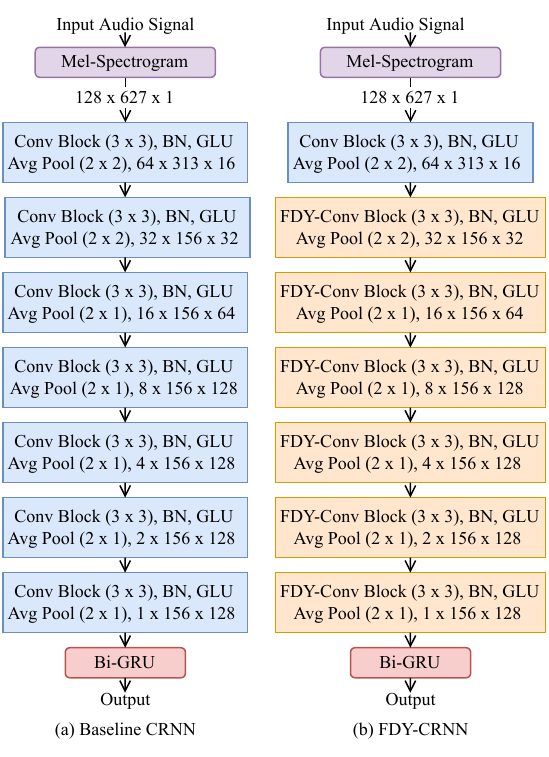}
\vspace*{-4mm}
\caption{\textit{Architecture of (a) CRNN (Baseline) (b) FDY-CRNN. when hop length is 256.}}
\label{fig:fdy}
\vspace*{-6mm}
\end{figure}

% One figure for the baseline model

% After the mapping, the dataset can be divided into three parts: 1. Original DESED dataset with DESED distinguish classes, 2. Original MAESTRO dataset with MAESTRO distinguish classes, 3. Original MAESTRO dataset with DESED and MAESTRO common classes.

\subsection{Domain generalization}
\label{ssec:dg}
In DCASE 2024 Challenge Task 4, we must train one model on heterogeneous datasets. As mentioned in Section~\ref{ssec:dataset}, the DESED and MAESTRO Real datasets come from different sources. Even if they share the same segment labels, their features can be quite different, confusing model training because domain information is not useful. domain generalization (DG)~\cite{dg} aims to address this problem by learning robust models against data distribution changes across domains, known as domain shift. The goal is to ensure that the trained model can generalize well to any domain by learning domain-invariant feature representations that remain discriminative across multiple domains. 

MixStyle~\cite{mixstyle} is a common DG method motivated by the observation that the visual domain is closely related to image style. Specifically, MixStyle mixes the feature statistics of two instances with a random convex weight to simulate new styles. However, unlike images where 2D convolution operates along spatial dimensions, in case of audio, 2D convolution operates on frequency and temporal information. Therefore, domain information may not be mainly distributed in channel statistics in audio as it is in MixStyle. In~\cite{freqmixstyle}, the authors analyzed the relationship between the domain and the statistics of each feature dimension showing that the frequency feature dimension carries more domain-relevant information than the channel dimension. This inspired us to adopt frequency-wise MixStyle for applying it to both internal CNN features and the mel-spectrogram before feeding data into the CNN. After integrating frequency-dynamic attention and MixStyle, we propose a new model, which is introduced in the next section. We note that MixStyle is not applied during testing.

We also explored several other DG methods including the residual normalization~\cite{resnorm}, and adapted them to the audio task. However, the experiments demonstrate that the freqwise mixstyle has straightforward improvement for heterogeneous training datasets.
\subsection{Network}
\label{ssec:network}
In this work, we also employed FDY-CRNN from \cite{fmsg1}, which uses frequency-adaptive kernels to enforce frequency dependency in 2D convolutions. In the baseline CRNN architecture shown in Figure 1(a), we replaced the standard 2D convolutional blocks with FDY-convolutional blocks, as illustrated in Figure 1(b). The CNN part consists of 7 blocks with the same number of filters as in the baseline. In the FDY-convolutional block, batch normalization and gated linear units are used.

\subsection{Independent loss function}

The model is trained using the binary cross-entropy (BCE) loss function on DESED real-world strongly, synthetic, and weakly labeled data, as well as on MAESTRO soft-labeled data. Mean squared error (MSE) is used for the mean-teacher~\cite{meanteacher,meanteacherbaseline1} pseudo-labeling loss component, which is applied to both weak and unlabeled data from DESED. When computing the loss for both components on a particular clip, we avoid computing the loss for the network outputs corresponding to classes that do not belong to the clip's original dataset. For example, in case of MAESTRO, we do not compute the loss for DESED output logits even for classes that have been cross-mapped as explained in the baseline. This is the main difference between our system and the baseline in loss computation.

\subsection{Pretrained model}
\label{ssec:pretrained}

% There are several pretrained models that can be considered for adoption. Recently, the audio teacher-student transformer demonstrated state-of-the-art performance in the DESED 2023 Task 4A. However, when we attempted to reproduce this model with heterogeneous training datasets, it didnot perform well on the MAESTRO dataset, showing a tendency to overfit on the DESED dataset. Therefore, w
We utilize the pretrained BEATs model, which has achieved state-of-the-art performance on AudioSet with a mean average precision (mAP) of 0.486. The BEATs is an iterative self-supervised framework for audio representation learning, using an acoustic tokenizer and a semi-supervised learning model. Unlike previous models, BEATs employs a self-distilled tokenizer to convert audio signals into discrete labels. We use it to construct frame-level embeddings of size 768, aligning with the recently released baseline approach.

\subsection{Data augmentation}

For this year's challenge, we used two data augmentation methods. We applied SpecAugment-style time-wise masking~\cite{specaug} to the features extracted by the pre-trained model and independently to the features extracted from the CNN encoder. This strategy, referred to as ``dropstep", helps improve model robustness by adding variety to the training data.

Additionally, we used the Mixup~\cite{mixup} strategy. This helps in linear interpolation and improves model robustness. Mixup is applied independently on the MAESTRO and DESED datasets.

\subsection{Curated set}
\label{ssec:curated}
To solve the mismatch between the synthetic valid dataset and the real-world test set, we further split the 3,470 clips of the strongly annotated AudioSet part to get the extra real valid dataset (373 clips). 

\subsection{Sound event box-based post-processing}
Existing systems~\cite{postprocessing} commonly predict sound presence confidence in short time frames. Then, thresholding produces binary frame-level presence decisions, with the extent of individual events determined by merging consecutive positive frames. In the previous challenge~\cite{fmsg1}, we used median filtering as post-processing. 

A recent study in~\cite{sebbs} shows that frame-level thresholding degrades the prediction of event extent by coupling it with the system’s sound presence confidence. Inspired by bounding box predictions in image object detection~\cite{yolo} SEBBs are one-dimensional bounding boxes defined by event onset time, event offset time, sound class, and confidence. They represent sound event candidates with a scalar confidence score. The final SED is derived by class-wise event-level thresholding of the SEBBs' confidences. SEBBs whose confidence exceeds the threshold are accepted as detections, while the rest are discarded. The threshold controls the sensitivity of systems. For high sensitivity/recall (few missed hits), a low detection threshold detects events even when the system's confidence is low. For high precision (few false alarms), a higher threshold detects only events with high confidence. With SEBBs, the sensitivity of a system can be controlled without impacting the detection of an event's onset and offset times, which was a problem with the previous frame-level thresholding approach.

We first tune the hyperparameters for the change-point-based predictor of Sound Event Bounding Boxes (cSEBBs)~\cite{sebbs} based on the strong validation dataset, then we use cSEBBs as the post-processing method in our system.

\section{Experimental setup}
\label{sec:experiments}
\subsection{Feature extraction}

All audio clips are resampled to a 16 kHz mono channel using Librosa. They are segmented with a window size of 2048 samples and a hop length of either 160 or 256 samples. A short-time Fourier transform is applied to extract spectrograms. Mel-filters are then used to create log-mel spectrograms spanning from 0 to 8 kHz. Clips shorter than 10 seconds are padded with silence if needed.
\begin{table*}[t]
\centering
\caption{Performance in PSDS and mPAUC of different single-systems on the DESED development set (D-PSDS), DESED public evaluation set (PE-PSDS), and MAESTRO evaluation set (mPAUC) including frequency-dynamic convolution (FDY), domain generalization (DG), and strong validation dataset (Strong Val). `AdaResNorm' and `ResNorm' stands for adaptive residual normalization and residual normalization, respectively. Based on the joint score, the sum of PE-PSDS and mPAUC, the systems with `*' are chosen for final submission.}
\vspace{2mm}
\label{tab:single}
\resizebox{\textwidth}{!}{%
\begin{tabular}{|c|c|c|c|c|c|c|c|c|c|}
\hline
\multicolumn{1}{|c|}{\textbf{System}} &
  \textbf{DG} &
  \textbf{Hop\_length} &
  \textbf{FDY} &
  \textbf{Strong Val} &
  \textbf{D-PSDS} &
  \textbf{PE-PSDS (raw)} &
  \textbf{PE-PSDS (cSEBBs)} &
  \textbf{mPAUC} &
  \textbf{Joint score}
  \\ \hline\hline
Baseline & - & 256 & - & - & 0.483 & 0.529 & - & 0.721 & 1.250 \\ \hline
S-1* & MixStyle & 256 & - & - & 0.506 &  0.587  & 0.629 & 0.737 & 1.366\\ \hline
S-2 & MixStyle & 256 & \ding{51} & - & \textbf{0.510} & 0.590 & 0.600 & \textbf{0.753} & 1.353 \\ \hline
S-3* & MixStyle inside FDY & 256 & \ding{51} & - & 0.503 & 0.590 & 0.634 & 0.737 & 1.371\\ \hline
S-4 & MixStyle & 256 & - & \ding{51} & 0.500 & 0.595 & 0.615 & 0.745 & 1.360\\ \hline
S-5 & AdaResNorm & 256 & - & - & 0.493 & 0.589 & 0.593  & 0.747 & 1.340\\ \hline
S-6 & ResNorm & 256 & - & - & 0.491 & 0.595 & 0.595 & 0.733 & 1.328\\ \hline
S-7* & MixStyle & 160 & - & - & 0.480 & 0.588 & \textbf{0.643} & 0.748 & \textbf{1.391}\\ \hline
S-8 & MixStyle inside FDY & 160 & \ding{51} &-& 0.485 & \textbf{0.599} & 0.629 & 0.737 & 1.366\\ \hline
\end{tabular}%
}
% \vspace{-5mm}
\end{table*}

\subsection{Training method}
For all experiments, a batch size of 60 was used, comprising the strong set, weak set, and unlabeled set, with batch size distribution: approximately 1/5 of the maestro dataset, 1/10 of the synth dataset, 1/10 of the synth+strong dataset, 1/5 of the weak dataset, and 2/5 of the unlabeled dataset. The training process included 50 epochs for warmup, a maximum of 300 epochs, and an epoch decay of 100. Gradient clipping was set at 5.0, the EMA factor for the mean teacher~\cite{meanteacher} was 0.999, validation was performed every 10 epochs, and the maximum weight for self-supervised loss was 2. The Adam optimizer was employed with a learning rate of 0.001. An exponential warmup was applied for the initial 50 epochs, and no early stopping was implemented during the training process.
\subsection{Evaluation metric}
This year, this task requires us to consider the PSDS~\cite{psds,tpsds} for evaluation. Event onset and offset times required for PSDS computation are only available for DESED data and classes, so PSDS is only evaluated on this fraction of the evaluation set. For MAESTRO, segment-based labels (one second) are provided, and we use the segment-based mean (macro-averaged) partial area under the ROC curve (mPAUC) as the primary metric, with a maximum FP-rate of 0.1. mPAUC is computed with respect to hard labels (threshold = 0.5) for the 11 classes listed. DESED and MAESTRO clips are anonymized and shuffled in the evaluation set to prevent manual domain identification.

\subsection{Ensemble}
\label{ssec:ensemble}
Ensemble modeling is a technique that leverages the strengths of multiple models to improve overall performance and enhance the generalization capability of a system. 
% By combining the predictions from different models, ensemble methods can effectively reduce individual model biases and errors, leading to more accurate and robust results. 
In our system, ensemble modeling plays a crucial role to improve system performance. These models work together to extract the best aspects from the highest-performing models. To generate final predictions, we aggregate the individual predictions from all the models and calculate their average. This approach ensures that every model contributes to the overall performance of the ensemble system.

\section{Results and Analysis}
\label{sec:results}
In this section, we first present the findings of the of the 3 single systems, followed by 1 ensemble system that we submitted.

% \usepackage{graphicx}

\subsection{Single-systems}
The PSDS measures the performance of systems on SED for the DESED subsets. In the DESED development set, the baseline system achieves a PSDS of 0.483. Systems S-1* (0.506), S-2 (0.510), S-3* (0.503), and S-4 (0.500) outperform the baseline, showing significance of  MixStyle to improve performance. Systems S-5 (0.493) and S-6 (0.491) show slight improvements, while S-7* (0.480) and S-8 (0.485) perform slightly worse, suggesting the hop length of 160 might be less effective. For the public evaluation set, while the baseline achieves 0.529 our systems S-1* (0.587), S-2 (0.590), S-3* (0.590), S-4 (0.595), and S-8 (0.599) outperform the baseline, with S-8 achieving the highest score, indicating the benefit of integrating MixStyle inside FDY.

The mPAUC is another key metric that is used for MAESTRO subset. The baseline system achieves an mPAUC of 0.721. System S-1* shows a significant improvement with 0.737, highlighting the positive impact of MixStyle on generalization. System S-2 achieves the highest mPAUC of 0.753, which indicates combining MixStyle with frequency-dynamic convolution significantly enhances performance. System S-3* also performs well with 0.737, reinforcing the benefits of domain generalization. Systems S-5 (0.747) and S-6 (0.733) show moderate improvements with adaptive normalization techniques, while S-7* (0.748) demonstrates excellent performance even with a shorter hop length. System S-8, integrating MixStyle inside FDY, shows consistent improvement with an mPAUC of 0.737.

When combining results from both PSDS and mPAUC, it becomes more evident that systems integrating MixStyle and frequency-dynamic convolution (FDY) outperform the baseline across different datasets. System S-8, for instance, achieves the highest public evaluation set PSDS (0.599) and a strong mPAUC (0.737), demonstrating its robustness and adaptability. Systems S-1* and S-3* also show balanced improvements across both metrics, making them reliable choices for the final submission.

We chose systems S-1*, S-3*, and S-7* for final submission due to their superior performance and generalization capabilities. System S-1* achieves high scores across all metrics, indicating robust generalization. System S-3* combines MixStyle with frequency-dynamic convolution, effectively enhancing domain generalization and event-level detection. System S-7* shows excellent performance with a shorter hop length, demonstrating the model’s adaptability to different configurations. These single-systems consistently outperform the baseline and showed significant improvements in terms of both the performance metrics, making them the best candidates for submission.

\subsection{Ensemble system}
We used all the single-systems except S-5 and S-6 for the ensemble system. After applying the ensemble method mentioned in the previous section, we achieved the highest development dataset PSDS of 0.520, a raw public evaluation dataset PSDS of 0.620, and an mPAUC of 0.762. After applying cSEBBs post-processing, the public evaluation dataset PSDS improved to 0.656, and thereby improved the joint score to {\bf 1.418}. We submitted this system as our only ensemble system.

\section{Conclusion}

This report presents SED systems by Fortemedia Singapore and the Joint Laboratory of Environmental Sound Sensing as a participation to DCASE 2024 Task 4. We addressed the challenge of recognizing overlapping events from different sources using a method that integrates bidirectional encoder representations from audio transformers and a convolutional recurrent neural network. Our key strategies included applying MixStyle to adapt multi-domains, using an independent learning framework for dataset-specific training loss, and employing sound event bounding boxes for post-processing. As a part of our submission we submitted 3 single-systems and 1 ensemble system. Our ensemble system achieved the highest public evaluation dataset PSDS of 0.656, and mPAUC of 0.762, thereby showed significant improvment over the challenge baseline.
\clearpage
% -------------------------------------------------------------------------
% Either list references using the bibliography style file IEEEtran.bst
\footnotesize
\bibliographystyle{IEEEtran}
\bibliography{refs}
%
% or list them by yourself
% \begin{thebibliography}{9}
% 
% \bibitem{dcase2016web}
%   \url{http://www.cs.tut.fi/sgn/arg/dcase2016/}.
%
% \bibitem{IEEEPDFSpec}
%   {PDF} specification for {IEEE} {X}plore$^{\textregistered}$,
%   \url{http://www.ieee.org/portal/cms_docs/pubs/confstandards/pdfs/IEEE-PDF-SpecV401.pdf}.
%
% \bibitem{PDFOpenSourceTools}
%   Creating high resolution {PDF} files for book production with 
%   open source tools, 
%   \url{http://www.grassbook.org/neteler/highres_pdf.html}.
%
% \bibitem{eWilliams1999}
% E. Williams, \emph{Fourier Acoustics: Sound Radiation and Nearfield Acoustic
%   Holography}. London, UK: Academic Press, 1999.
% 
% \bibitem{ieeecopyright}
%   \url{http://www.ieee.org/web/publications/rights/copyrightmain.html}.
%
% \bibitem{cJones2003}
% C. Jones, A. Smith, and E. Roberts, ``A sample paper in conference
%   proceedings,'' in \emph{Proc. IEEE ICASSP}, vol. II, 2003, pp. 803--806.
% 
% \bibitem{aSmith2000}
% A. Smith, C. Jones, and E. Roberts, ``A sample paper in journals,'' 
%   \emph{IEEE Trans. Signal Process.}, vol. 62, pp. 291--294, Jan. 2000.
% 
% \end{thebibliography}

\end{sloppy}
\end{document}